\documentclass[aps, a4paper,superscriptaddress, nofootinbib, showpacs, twocolumn, 10pt]{revtex4}
\usepackage{amsmath,amssymb,bm,natbib}
\usepackage{epsfig}
\usepackage{graphicx}
\usepackage{slashed}

\newcommand{\beq}{\begin{equation}}
\newcommand{\eeq}{\end{equation}}
\newcommand{\bea}{\begin{eqnarray}}
\newcommand{\eea}{\end{eqnarray}}
\newcommand{\ba}{\begin{align}}
\newcommand{\ea}{\end{align}}
\newcommand{\bfig}{\begin{figure}}
\newcommand{\efig}{\end{figure}}

\newcommand{\D}{\displaystyle}

\newcommand{\gev}{\, \text{GeV}}
\newcommand{\mev}{\, \text{MeV}}

\newcommand{\tin}{t_{\rm in}}
\newcommand{\la}{\langle}
\newcommand{\ra}{\rangle}

\newcommand{\omnes}{{\cal{O}}}

\begin{document}

\vspace{1cm}

\title{Spacelike pion form factor from analytic continuation and the onset of perturbative QCD}
\author{B.Ananthanarayan}
\affiliation{Centre for High Energy Physics,
Indian Institute of Science, Bangalore 560 012, India}
\author{Irinel Caprini}
\affiliation{Horia Hulubei National Institute for Physics and Nuclear Engineering,\\
P.O.Box MG-6, 077125 Magurele, Romania}
\author{I. Sentitemsu Imsong}
\affiliation{Centre for High Energy Physics,
Indian Institute of Science, Bangalore 560 012, India}

~\vspace{0.5cm}

\begin{abstract}
The factorization theorem for exclusive processes in 
perturbative QCD predicts the behavior
of the pion electromagnetic form factor 
$F(t)$ at asymptotic spacelike momenta $t(=-Q^2)<0$.  
We address the question of the onset energy  using a suitable
mathematical framework of analytic continuation, which uses as input 
the phase of the form factor below the  
first inelastic threshold, known with 
great precision through the Fermi-Watson theorem 
from  $\pi\pi$ elastic scattering,  and the modulus measured from threshold  
up to 3 GeV by the BaBar Collaboration. The method  leads to almost model-independent 
upper and lower bounds on the spacelike form factor.
Further inclusion of the value of the charge radius and 
the experimental value at $-2.45 \gev^2$  measured  at JLab considerably increases the strength of the bounds in the region $ Q^2 \lesssim 10 \gev^2$, excluding the onset of the 
asymptotic perturbative QCD regime for $Q^2< 7\gev^2$. 
We  also compare the bounds  with available experimental data  
and with several theoretical models proposed 
for the low and intermediate spacelike region.
\end{abstract}

\pacs{11.55.Fv, 13.40.Gp, 25.80.Dj}
\maketitle
\section{Introduction}
\label{sec:intro}

The high-energy behavior along the spacelike axis of the pion electromagnetic form factor is 
predicted by factorization in perturbative QCD 
\cite{Farrar:1979aw,Lepage:1979zb,Efremov:1979qk,pQCD3}, which to 
leading order (LO) gives\footnote{In our convention the form factor $F(t)$ is a function of 
the squared momentum transfer $t=q^2$, and is a real analytic function in the complex $t$-plane 
cut along  $t\geq 4 M_\pi^2$. In this notation the spacelike axis is defined by $t(=-Q^2)<0$.}   
 \beq\label{eq:qcd1}
  F^{\rm LO}_{\rm pert}(-Q^2)= \frac{8 \pi f_\pi^2\alpha_s(\mu^2 )}{Q^2},  
 \eeq
where $f_\pi= 130.4 \mev$ is the pion decay constant and $\alpha_s(\mu^2)$ the  strong  coupling at 
the renormalization scale $\mu^2$. Next-to-leading-order (NLO)  perturbative  corrections have been 
calculated in 
\cite{Field:1998,Dittes:1981,Khalmuradov:18985,Braaten:1987,Melic:1998qr,Melic:1999mx,Li:2011}, using 
various renormalization schemes and  pion 
distribution amplitudes (DAs). In particular, the result in the ${\overline{\rm MS}}$-renormalization 
scheme with asymptotic DAs reads \cite{Melic:1999mx}
\beq\label{eq:qcd2}
F^{\rm NLO}_{\rm pert}(-Q^2)= \frac{8 f_\pi^2\alpha_s^2(\mu^2)}{Q^2}\,\left[\frac{\beta_0}{4}\left( \ln \frac{\mu^2}{Q^2}+\frac{14}{3}\right) - 3.92\right],
 \eeq
where $\beta_0=11-2 n_f/3$ is the first
coefficient in the perturbative expansion of the $\beta$-function, 
$n_f$ being the number of active flavors.
The ambiguities that affect the perturbative QCD predictions have been investigated in many papers, 
where, in particular, the dependence on the renormalization scale $\mu^2$ and  various prescriptions for scale 
setting have been discussed \cite{Brodsky:1998,Melic:1999mx}. 

It has been known for a long time that in the case of the pion form factor the asymptotic regime 
sets in quite slowly, due to the complexity of soft, nonperturbative processes in QCD in the intermediate
$Q^2$ region.  Many nonperturbative approaches have been proposed for the study of the form factor, 
including QCD sum rules \cite{Ioffe:1982}, quark-hadron local duality  
\cite{Nesterenko:1982,Radyushkin:1995, Radyushkin:2001,Braguta:2007fj}, 
extended vector meson dominance \cite{Dominguez},
light-cone sum rules 
\cite{Braun:1994,Braun:1999uj,Bijnens:2002}, sum rules with nonlocal condensates 
\cite{Bakulev:1991ps,Bakulev:2004cu,Bakulev:2009ib}, and anti-de Sitter (AdS)/QCD models \cite{Grigoryan:2007,Brodsky:2007hb}.  
The scale of the onset in the presence of Sudakov corrections \cite{Gousset}
and large $N_c$ Regge approaches \cite{RuizArriola} suggests that it can be very large.
However, constructing a fully valid model to describe the form factor at intermediate
energies in fundamental QCD still remains a major theoretical challenge. 

On the experimental side, several  measurements of the spacelike form factor at various energies are 
available (see Refs. 
\cite{Brown:1973wr,Bebek:1974iz,Bebek:1976ww,Ackermann:1977rp,Bebek,Brauel:1979zk,Amendolia:1986wj,
Volmer:2000ek,Tadevosyan,Horn,Huber}), the most precise being the recent results of 
the JLab Collaboration \cite{Horn, Huber} 
with the highest measurement being at $Q^2=2.45 \gev^2$. 
The experimental determination
of $F$ at larger values of $Q^2$ is  difficult due to the lack of a free pion target
and requires the use of pion electroproduction from a nucleon target.
This stems from a virtual photon coupling to a pion in the cloud
surrounding the nucleon. In this regard, there are uncertainties
associated with the off shellness of the struck pion and the
consequent extrapolation to the physical pion mass pole, which
leads to uncertainties in the extraction of the value of the
form factor at $t=0$, and possible contributions from nonresonant
backgrounds. The lack of reliable experimental data in the higher $Q^2$ region 
is a major obstacle to confirm or discard the number of theoretical models already available. In particular,
it remains an open question as to at what value of $Q^2$ do the nonperturbative contributions
become negligible so that the perturbative QCD description of the form factor becomes reliable.

In the recent past, the knowledge of the form factor has improved considerably on the timelike axis. 
The phase of the form factor in the elastic region of the unitarity cut (below the first inelastic threshold 
associated with the $\pi\omega$ state) is equal, by the final state interaction
theorem also known as the Fermi-Watson 
theorem, to the $P$-wave phase shift of the  elastic $\pi\pi$ scattering amplitude, which has been calculated 
recently with precision using Roy equations  \cite{ACGL,CGL,GarciaMartin:2011cn,CapriniRegge:2012}.  
The modulus has been measured from the cross section of $e^+e^-\to\pi^+\pi^-$ by several groups in the 
past, and more recently to high accuracy by BaBar \cite{BABAR} and
KLOE \cite{KLOE1,KLOE2} Collaborations. In particular, the high statistics measurement by BaBar 
yields the  modulus at high precision up to an energy of 3 GeV. Therefore, an interesting 
possibility is to find constraints on the spacelike form factor by the analytic extrapolation of the accurate
timelike information. 

In the present paper we address precisely this problem. Namely, our aim 
is to perform an analytic continuation of the form factor from the timelike to the spacelike 
region using in a most conservative way the available information on the phase and modulus on 
the unitarity cut, and also spacelike information.
The main result of the method will consist of rather stringent upper and lower bounds at values
of spacelike momenta, which provide a criterion for finding a lower limit
for the onset of the QCD perturbative behavior.

We start by briefly discussing, in  Sec.\ref{sec:continuation}, several methods of analysis based on 
the analytic properties of the form factor investigated in the literature 
\cite{DoNa,Belicka:2011ui,TrYn1,GuPi,PiPo,Geshkenbein,BuLe,LaSt,HeLa,Colangelo:2004,Masjuan,Leutwyler:2002}.  
We then present our method, which uses the available input in a very conservative way.  As we will explain, 
the price to be paid is that we are able to derive only upper and lower bounds on the spacelike form factor.  
The extremal problem leading to these bounds is formulated accordingly. 

 In  Sec.\ref{sec:method} we present the solution of the extremal problem formulated in the previous section.
We use a mathematical technique applied for the first time in \cite{IC} and reviewed more recently in 
\cite{Abbas:2010EPJA}, which has been exploited also for the investigation of the low-energy shape 
parameters and the location of the zeros of several hadronic form factors 
\cite{Abbas:2009dz, Abbas:2010EPJA, Abbas:2010ns, Ananthanarayan:2011xt, Ananthanarayan:2011uc}. We emphasize that the 
solution of the problem is exact, so within the adopted assumptions the bounds are optimal.

In Sec.\ref{sec:inputs} we gather all the theoretical and phenomenological inputs that go into our
calculation, and in Sec.\ref{sec:results} we  present the bounds on the form factor in the spacelike 
region derived using our formalism. In this section we also 
 compare our findings with both experimental and theoretical
results available, being able, in particular, to investigate  the onset of the
asymptotic regime of  perturbative QCD and to check the validity of various nonperturbative models 
proposed in the literature.   Finally in Sec.\ref{sec:conclusions}, we present 
a brief discussion on the implication of our results and draw our conclusions.

\section{Analytic continuation}\label{sec:continuation}
The standard dispersion relation in terms of the imaginary part on the form factor along the unitarity 
cut is the simplest way to perform the analytic continuation from the timelike to the spacelike region. 
This method was applied for a discussion of perturbative QCD in \cite{DoNa}, and more recently  in 
\cite{Belicka:2011ui}. As the imaginary part of the form factor is not directly measurable, the 
method requires the knowledge of both the phase and the modulus, and is influenced by errors coming 
from the relatively poor determination of the modulus at low energies, and the lack of knowledge of 
the phase above the elastic region.

Alternative dispersive analyses applied so far are based on representations either in terms of 
the phase, the so-called  Omn\`es representation \cite{GuPi, PiPo,TrYn1}, or in terms of  the modulus 
\cite{Geshkenbein}. Both approaches require the knowledge of zeros of the form factor, and are plagued, 
respectively,  by uncertainties related to the unknown phase in the inelastic region, and the 
uncertainties of the modulus at low energies.  Various analytic representations or expansions in 
terms of suitable sets of functions were also used in \cite{LaSt}-\cite{Leutwyler:2002} for data 
analysis  and the analytic extrapolation of the form factor. 

Our approach aims to use as input in a most conservative way the  precise information available 
on the unitarity cut. First, we consider  the relation
\beq\label{eq:watson}
{\rm Arg} [F(t+i\epsilon)]=\delta_1^1(t), \quad\quad  4 M_\pi^2 \leq t \leq \tin
\eeq
where $\delta_1^1(t)$ is the phase shift of the $P$ wave of $\pi\pi$ elastic scattering. We denoted by
$\tin$  the upper limit of the elastic region,  which can be taken as $\tin=(M_\pi+M_\omega)^2$ 
since the first important inelastic threshold in the unitarity relation for the pion form factor 
is due to the $\omega\pi$ pair. 

 Under weak assumptions, the asymptotic behavior (\ref{eq:qcd1}) along the spacelike axis 
implies a decrease like $|F(t)| \sim 1/t$ also on the timelike axis.\footnote{We quote a 
rigorous result given in \cite{Cornille:1975}, which states that, if an analytic function $F(t)$ 
has a bounded phase for $t\to -\infty$ and  $t\to \infty$, the ratio $|F(-t)/F(t)|$ tends 
asymptotically to 1, at least in an averaged sense.} Therefore, using the recent experimental
data on the modulus up to $\sqrt{t}=3\, \gev$ \cite{BABAR}, supplemented with conservative 
assumptions above this energy,  we can obtain a rather accurate estimate of an integral of 
modulus  squared from $\tin$ to infinity. More precisely, we assume the following condition,
 \beq\label{eq:L2}
 \D\frac{1}{\pi} \int_{\tin}^{\infty} dt \rho(t) |F(t)|^2 = I,
 \eeq
 where $\rho(t)$ is a suitable positive-definite weight, for which the integral converges,  
and the number $I$  can be estimated with sufficient precision.   The optimal procedure is to vary $\rho(t)$ over 
a suitable admissible class and take the best result. In principle, a large 
class of positive weights, leading to a convergent integral for $|F(t)|$ compatible with the asymptotic 
behavior  (\ref{eq:qcd1}) of the pion form factor, can be adopted. A suitable choice is
\beq \label{eq:rhogeneric0}
\rho_{b,c}(t, Q_0^2) = \frac{t^b}{(t+Q_0^2)^c}, \quad \quad 
\eeq
with  the parameter $Q_0^2>0$ and $b, c$ in the range $0\leq c-b \leq 2$. In most of our 
calculations we shall choose the simpler form 
\beq \label{eq:rhogeneric}
\rho_a(t) = \frac{1}{t^a}, \quad \quad 
\eeq
with $0\leq a \leq 2$.   The constraint (\ref{eq:L2}) and the choice of the optimal 
weight will be discussed in more detail later. 

 Additional information inside the analyticity domain can be implemented exactly.  
In practice we shall use the input
\beq\label{eq:taylor}
F(0)=1, \quad \quad \quad  F'(0)=\frac{1}{6} \la r_\pi^2 \ra,
\eeq
with the charge radius $\la r_\pi^2 \ra$ varied within reasonable limits, and the values 
of the form factor at some spacelike values
\beq\label{eq:Huber}
F(t_n)= F_{n} \pm \epsilon_n,  \quad \quad t_n<0,
\eeq
where $F_n$ and $\epsilon_n$ represent the central value and the  experimental uncertainty, 
known from the most precise experiments \cite{Horn, Huber}.

The relations (\ref{eq:watson}),  (\ref{eq:L2}), (\ref{eq:taylor}) and (\ref{eq:Huber}) 
define a specified class of real analytic functions in the $t$-plane cut for $t>4 M_\pi^2$. 
The problem is to derive rigorous upper and lower bounds on $F(t)$ in the region $t<0$, for 
functions $F(t)$ belonging to this class. This extremal problem will be solved exactly in the next section.

\section{Solution of the mathematical problem}\label{sec:method}
For solving the problem formulated in the previous section,  we apply a standard mathematical 
method  \cite{IC,Abbas:2010EPJA}.  We first define the Omn\`{e}s function
\beq	\label{eq:omnes}
 \omnes(t) = \exp \left(\D\frac {t} {\pi} \int^{\infty}_{4 M_\pi^2} dt' 
\D\frac{\delta (t^\prime)} {t^\prime (t^\prime -t)}\right),
\eeq
where $\delta(t)=\delta_1^1(t)$   for 
$t\le \tin$, and is an arbitrary function, sufficiently  smooth ({\em i.e.,}
Lipschitz continuous) for $t>\tin$. As shown in  \cite{Abbas:2010EPJA}, the results do not depend on the 
choice of the function  $\delta(t)$ for $t>\tin$.
The crucial remark is that the function $h(t)$ defined by
\beq\label{eq:h}
F(t)=\omnes(t) h(t)
\eeq
is analytic in the $t$-plane cut only for $t>\tin$. Furthermore, equality (\ref{eq:L2}) implies that $h(t)$ 
satisfies the condition
\beq\label{eq:hL2}
\D\frac{1}{\pi} \int_{\tin}^{\infty} dt\, 
\rho(t) |\omnes(t)|^2 |h(t)|^2 = I.
\eeq
 This relation can be written in a canonical form, if we perform the conformal transformation
\beq\label{eq:ztin}
\tilde z(t) = \frac{\sqrt{\tin} - \sqrt {\tin -t}} {\sqrt{\tin} + \sqrt {\tin -t}}\,,
\eeq
which maps the complex $t$-plane cut for $t>\tin$  onto the unit disk $|z|<1$ in the $z$ plane defined 
by $z\equiv\tilde z(t)$,  and define a function $g(z)$ by
\beq\label{eq:gF}
 g(z) = w(z)\, \omega(z) \,F(\tilde t(z)) \,[\omnes(\tilde t(z)) ]^{-1}, 
\eeq 
where $\tilde t(z)$ is the inverse of $z = \tilde z(t)$, for $\tilde z(t)$ as defined in
(\ref{eq:ztin}), and  $w(z)$ and $\omega(z)$ are calculable outer functions, {\it i.e.,} functions 
analytic and without zeros for $|z|<1$, defined in terms of their modulus on the boundary, 
related to  $\sqrt{\rho(t)}$ and  $|\omnes(t)|$,  respectively.

It follows from (\ref{eq:h}) that the product $F(\tilde t(z)) \,[\omnes(\tilde t(z)) ]^{-1}$ appearing in
(\ref{eq:gF}) is equal to the function $h(\tilde t(z))$ , 
which is analytic in  $|z|<1$. Therefore, the function $g(z)$ defined in (\ref{eq:gF}) is analytic in   $|z|<1$.

The outer functions corresponding to the weight functions (\ref{eq:rhogeneric0}) 
and (\ref{eq:rhogeneric}) can be written in an analytic closed form in the $z$ variable as
\beq\label{eq:outerfinal0}
w_{b,c}(z,Q_0^2)= (2\sqrt{t_{\rm in}})^{1+b-c}\frac{(1-z)^{1/2}} {(1+z)^{3/2-c+b}}\frac{(1+\tilde z(-Q_0^2))^c}{(1-z \tilde z(-Q_0^2))^c} ,
\eeq 
and, respectively,
\beq\label{eq:outerfinal}
w_a(z)= (2 \sqrt{t_{\rm in}})^{1-a}\,\frac{(1-z)^{1/2}} {(1+z)^{3/2-a}}.
\eeq
For the outer function $\omega$  we shall use an integral representation in terms of its modulus on the 
cut $t>\tin$, which can be written as \cite{IC,Abbas:2010EPJA}
\beq\label{eq:omega}
 \omega(z) =  \exp \left(\D\frac {\sqrt {\tin - \tilde t(z)}} {\pi} \int^{\infty}_{\tin}  \D\frac {\ln |\omnes(t^\prime)|\, {\rm d}t^\prime}
 {\sqrt {t^\prime - \tin} (t^\prime -\tilde t(z))} \right).
\eeq 

 Using the definition of the functions $w(z)$ and $\omega(z)$, it is easy to see that 
(\ref{eq:hL2}) can be written in terms of the function $g(z)$ defined in (\ref{eq:gF}) as
\beq\label{eq:gI1}
\frac{1}{2 \pi} \int^{2\pi}_{0} {\rm d} \theta |g(\zeta)|^2 = I, \quad\quad \zeta= {\rm e}^{i\theta}.
\eeq
We further note that  (\ref{eq:ztin}) implies that the origin $t=0$ of the $t$ plane is
mapped onto the origin $z=0$ of the $z$ plane. 
Therefore, from  (\ref{eq:gF}) it follows that each coefficient $g_k \in R$ of the expansion
\begin{equation}\label{eq:gz}
g(z)=g_0+ g_1 z+ g_2 z^2 +  g_3 z^3+\dots
\end{equation}
is expressed in terms of the coefficients of order lower or equal to $k$,  of the Taylor series 
expansion of the form factor at $t=0$. 
Moreover, the values $F(t_n)$ of the form factor at a set of real points
$t_n<0,\, n=1,2,..., N$ lead to the values
\beq\label{eq:xin}
g(z_n)=w(z_n)\, \omega(z_n) \,F(t_n) \,[\omnes(t_n) ]^{-1}, \quad z_n=\tilde z(t_n).
\eeq
 Then the $L^2$ norm condition (\ref{eq:gI1}) implies the determinantal inequality 
(for a proof and older 
references see \cite{Abbas:2010EPJA}):
\beq\label{eq:det}
\left|
\begin{array}{c c c c c c}
\bar{I} & \bar{\xi}_{1} & \bar{\xi}_{2} & \cdots & \bar{\xi}_{N}\\	
	\bar{\xi}_{1} & \D \frac{z^{2K}_{1}}{1-z^{2}_1} & \D
\frac{(z_1z_2)^K}{1-z_1z_2} & \cdots & \D \frac{(z_1z_N)^K}{1-z_1z_N} \\
	\bar{\xi}_{2} & \D \frac{(z_1 z_2)^{K}}{1-z_1 z_2} & 
\D \frac{(z_2)^{2K}}{1-z_2^2} &  \cdots & \D \frac{(z_2z_N)^K}{1-z_2z_N} \\
	\vdots & \vdots & \vdots & \vdots &  \vdots \\
	\bar{\xi}_N & \D \frac{(z_1 z_N)^K}{1-z_1 z_N} & 
\D \frac{(z_2 z_N)^K}{1-z_2 z_N} & \cdots & \D \frac{z_N^{2K}}{1-z_N^2} \\
	\end{array}\right| \ge 0,
\eeq
where 
\beq\label{eq:barxi}
 \bar{I} = I - \sum_{k = 0}^{K-1} g_k^2, \quad  \quad \bar{\xi}_n = g(z_n) - \sum_{k=0}^{K-1}g_k z_n^k,
\eeq
$K\ge 1$ denoting the number of Taylor coefficients from (\ref{eq:gz}) included as input.

 The inequality (\ref{eq:det}) leads to rigorous bounds on the value of the 
form factor at one spacelike point using input values at other spacelike points.
We can implement also a number $K$ of derivatives at $t=0$, in particular, the normalization 
and the charge radius from (\ref{eq:taylor}).  The derivation of the upper and lower bounds 
amounts  to solving simple quadratic equations for the quantities $g(z_n)$, related to the 
values $F(t_n)$ of the form factor through the relation (\ref{eq:xin}).  As shown in \cite{Abbas:2010EPJA}, 
the inequality (\ref{eq:det}) holds also if the equality sign in (\ref{eq:L2}) is replaced by the less 
than or equal sign. Furthermore, as argued in \cite{Abbas:2010EPJA}, the  bounds  depend in a monotonic way 
on the value of the quantity $I$  appearing in  (\ref{eq:gI1}), becoming weaker when this 
value is increased.

\section{Input}\label{sec:inputs}
 The phase shift  $\delta_1^1(t)$ was determined recently with high precision from Roy equations applied 
to the $\pi\pi$ elastic amplitude  in \cite{ACGL,CGL, GarciaMartin:2011cn,CapriniRegge:2012}. 
We use as phenomenological input the phase parametrized in \cite{GarciaMartin:2011cn} by
\beq\label{eq:delta11}
{\rm cot}\delta_1^1(t)=\frac{\sqrt{t}}{2k^3}(M_\rho^2 - t) \!\!\left(\frac{2 M_\pi^3}{M_\rho^2 \sqrt{t}} + B_0 + 
      B_1 \frac{\sqrt{t} -\sqrt{t_{0} - t}}{\sqrt{t} + \sqrt{t_{0} - t}}\!\!\right)\!, 
\eeq
where $k=\sqrt{t/4 -M_\pi^2}$, $\sqrt{t_0}=1.05\, \gev$,  $B_0=1.043\pm 0.011$, $B_1=0.19 \pm 0.05$ and $M_\rho= 773.6 \pm 0.9\, \mev$.  We assume isospin symmetry and take for $M_\pi$ the mass of the charged pion.

The function  $\delta_1^1$ obtained from (\ref{eq:delta11}) with the central values of the parameters 
is  practically identical with the phase shift  obtained in \cite{ACGL}  from Roy equations for 
$\sqrt{t}\leq 0.8\,\gev$, and its uncertainty in the whole elastic region $t<\tin$ is very small.

Above $\tin=(0.917\, \gev)^2$  we use in (\ref{eq:omnes}) a continuous function $\delta(t)$, 
which approaches asymptotically $\pi$. 
As shown in \cite{Abbas:2010EPJA}, if this function is sufficiently smooth (more exactly, 
Lipschitz continuous), the dependence on $\delta(t)$ of the functions $\omnes$ and $\omega$,  
defined in (\ref{eq:omnes}) and (\ref{eq:omega}), respectively, exactly compensate  each other, leading to 
results fully independent of the unknown phase in the inelastic region.

For the calculation of the integral defined in (\ref{eq:L2}) we
have used the BaBar data \cite{BABAR} 
from $\sqrt{\tin}=0.917 \gev$ up to $\sqrt{t}=3\, \gev$,  and 
have taken a constant value for the modulus 
in the range $3\, \gev \leq \sqrt{t} \leq 20 \gev$,  continued with a $1/t$ decrease above 20 GeV. 
This model is expected to overestimate the true value of the integral: indeed, we take up to 20 GeV the 
modulus equal to 0.066, {\em i.e.,} the central BaBar value at 3 GeV, while the perturbative QCD 
expression (\ref{eq:qcd1}), continued to the timelike axis,  predicts a much lower modulus, equal 
to 0.011 at 3 GeV and to 0.000 16 at 20 GeV. According to the above discussion,  
a larger value of $I$ leads to weaker bounds. Therefore, if we use 
an overestimate of  the integral, we weaken the bounds which nevertheless remain valid. 
This makes our procedure very robust. 

As we mentioned, in our analysis we shall work with weights of the form (\ref{eq:rhogeneric}). 
The values of $I$ corresponding to several choices of the parameter $a$ are given in Table \ref{table:Ia}, 
where the uncertainties are due to the BaBar experimental errors.  As expected, the most sensitive to 
the uncertainty of the high-energy behavior of $|F(t)|$ is the integral $I$ for $a=0$, because in this 
case the weight is constant. We checked, for instance, that if we use in  the integral above 20 GeV
the modulus $|F(t)|$ equal to 0.000 16 instead of 0.066,  we obtain for $I$   a value smaller by 
about 24\% for $a=0$, while the change in $I$ is of only 3\% for $a=1/2$ , and for $a=1$ and $a=2$ 
the values of $I$ remain practically unchanged. 

We complete the specification of our input by giving the range \cite{Masjuan,Colangelo:2004}
\beq\label{eq:r2}
  \la r_\pi^2 \ra =0.43 \pm 0.01 \,\mbox{fm}^2
\eeq
adopted for the charge radius, and the spacelike datum \cite{Horn,Huber}
\beq\label{eq:Huber2}
F( -2.45 \gev^2)= 0.167 \pm 0.010_{-0.007}^{+0.013}.
\eeq

We used the spacelike datum which gives the most stringent constraints. 
In the present case this is the point that is closest to the higher energies of interest.
In contrast, in the case of the shape parameters, the spacelike
data closest to the origin provide the most stringent 
constraints \cite{Ananthanarayan:2011xt}. 
Of course, more spacelike data may be included as input but, as discussed in
\cite{Ananthanarayan:2011xt}, when the uncertainties are taken into
account the improvement is not significant.

\begin{table}
\begin{center}
\caption{Values of the integral $I$ defined in (\ref{eq:L2}) corresponding  to different weight 
functions defined in  (\ref{eq:rhogeneric}). }\vspace{0.1cm}
\label{table:Ia}
\renewcommand{\tabcolsep}{1.5pc} 
\renewcommand{\arraystretch}{1.1} 
\begin{tabular}{cc}\hline
$a$  & $I$  \\\hline 
$0$ 			&  $1.788 \pm 0.039 $   \\
$1/2$ 	&  $0.687 \pm  0.028$\\
$1$ 		&  $0.578 \pm 0.022 $ \\ 
$2$ 		& $ 0.523 \pm 0.017 $ \\  
\noalign{\smallskip}\hline
\end{tabular}
\end{center}
\end{table}

\begin{figure}[htb]
 \vspace{0.35cm}
\begin{center}
  \includegraphics[width = 8.cm]{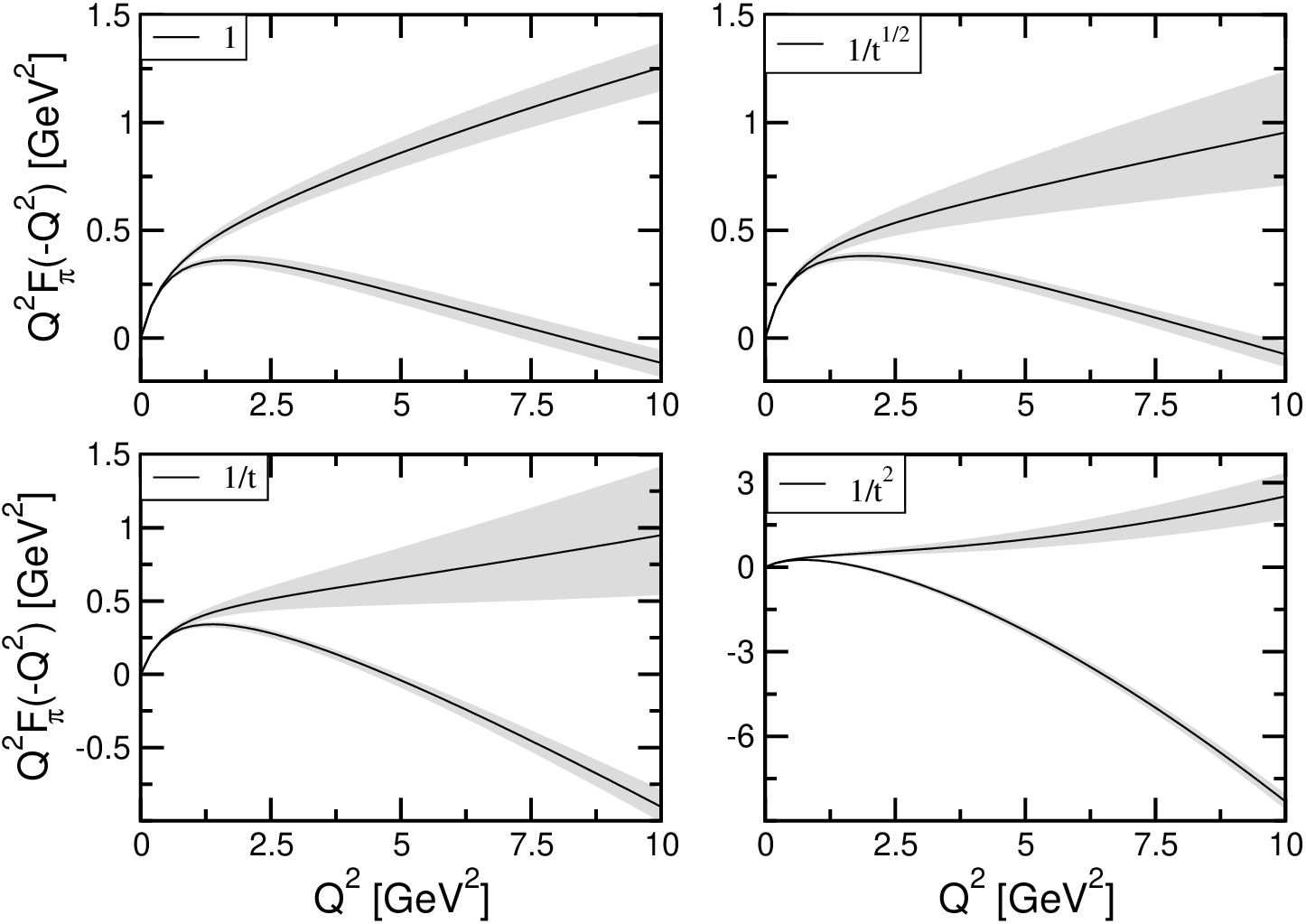}
 \caption{Upper and lower bounds obtained with several weight functions  $\rho_a(t)=1/t^a$. 
No spacelike information is used as input.  The black lines correspond to the central
values of the input and the grey bands indicate the errors. }
 \label{fig:fig2}
 	\end{center}
 \end{figure}

\begin{figure}[htb]
 \vspace{0.35cm}
\begin{center}
  \includegraphics[width = 8.cm]{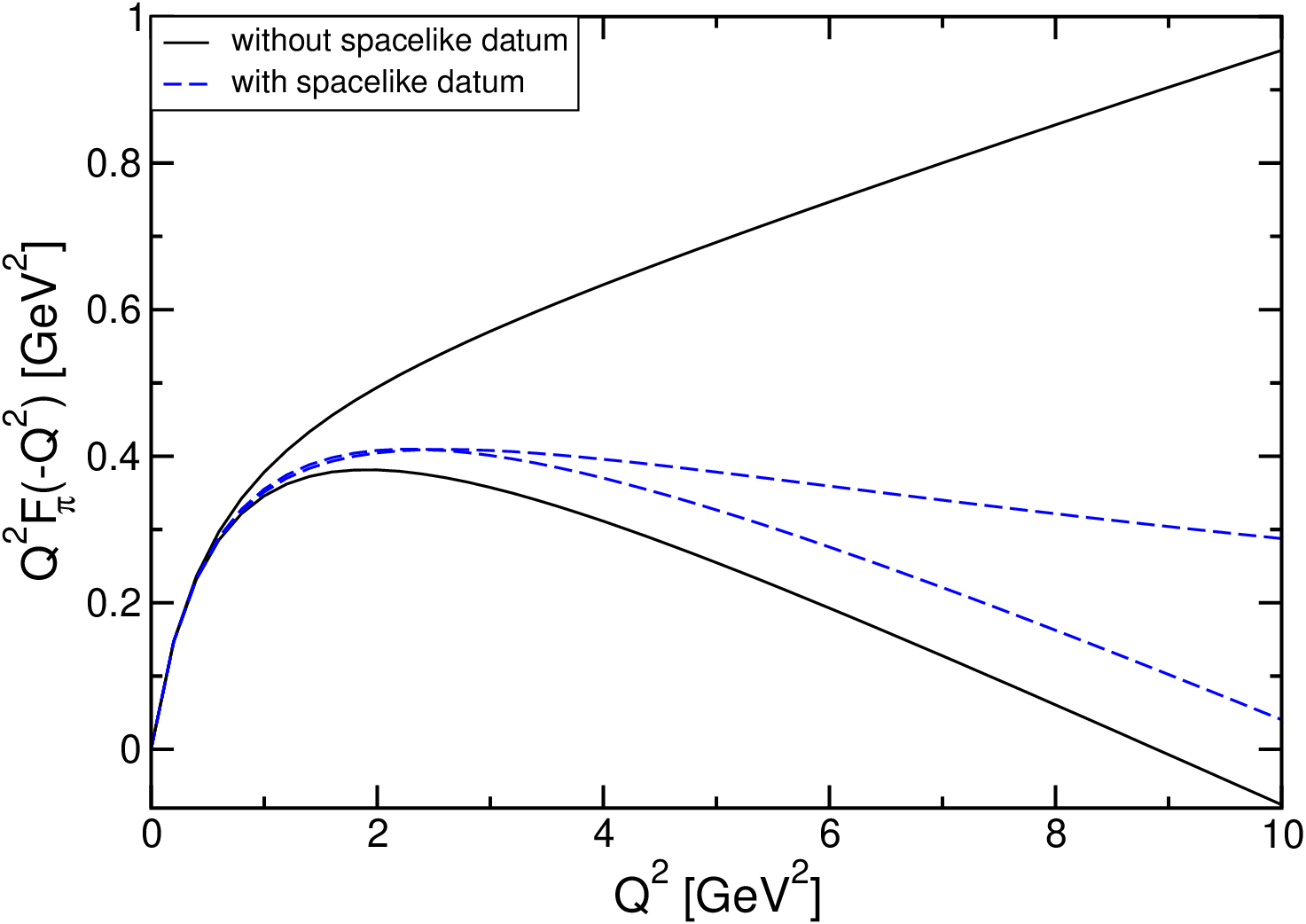}
 \caption{Effect of the additional spacelike input (\ref{eq:Huber2}) on the bounds obtained 
with the weight $\rho_{1/2}(t)$. Central values of all the input parameters are used.}
 \label{fig:fig4}
 	\end{center}
 \end{figure}

\bigskip

\begin{figure}[htb]
 \vspace{0.35cm}
\begin{center}
  \includegraphics[width = 8.cm]{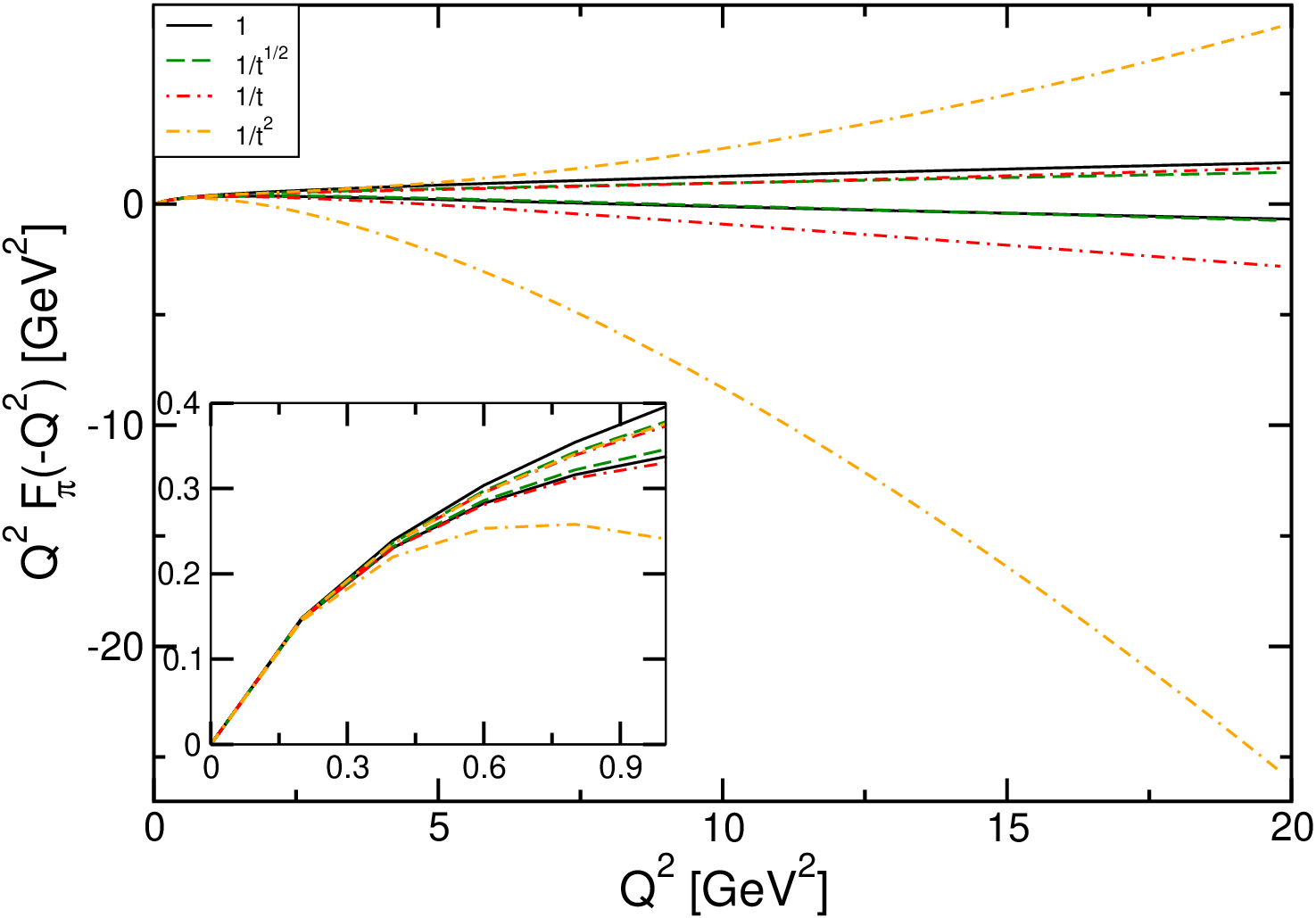}
 \caption{Upper and lower bounds obtained with various weights $\rho_a(t)=1/t^a$,
for the central values of the input parameters. The inset shows the bounds up to 1 GeV$^2$.}
 \label{fig:weights}
 	\end{center}
 \end{figure}

\section{Results}\label{sec:results}
For illustration we show first, in  Fig. \ref{fig:fig2}, the upper and lower bounds on the 
product $Q^2 \,F(-Q^2)$ in the range $0\leq Q^2 \leq 10\,\gev^2$, obtained with  different 
weight functions, using as input the relations (\ref{eq:watson}), (\ref{eq:L2}) and 
(\ref{eq:taylor}), with no information from the spacelike axis. 
The solid lines are the bounds for the central values of the input, and the grey bands 
indicate the uncertainty on the corresponding bounds, obtained by adding in quadrature the 
uncertainties due to the variation of the phase given in (\ref{eq:delta11}), the charge radius 
$\langle r_\pi^2 \rangle$ given in (\ref{eq:r2}), and  the integral $I$ from Table \ref{table:Ia}. 

 The results obtained with different weights are rather similar, except for the weight $\rho_2(t)$, 
which gives much weaker bounds at high $Q^2$. This feature is actually expected: since this weight 
decreases rapidly, the class of functions satisfying (\ref{eq:L2}) is less constrained at high 
energies, even functions increasing  asymptotically being accepted. We note that in the so-called 
unitarity  bounds approach, applied  to the pion form factor in 
\cite{IC, Ananthanarayan:2011xt}, an inequality 
similar to (\ref{eq:L2}) is derived  from  a dispersion relation  for a physical observable, 
like a polarization function calculated in perturbative QCD  
or the hadronic contribution to the muon anomaly, 
using in addition unitarity for the corresponding spectral function. In this approach the effective 
weight $\rho(t)$ is fixed and cannot be chosen at will,  decreasing actually like $1/t^2$ in all 
the cases investigated.  Therefore, as discussed in  \cite{Ananthanarayan:2011xt}, the approach of 
unitarity bounds, which is useful when no information on the modulus is available,  
would give weak bounds on the pion form factor at higher energies on the spacelike axis. 
The recent high statistic measurements of the modulus up to rather high energies on the 
timelike axis \cite{BABAR} allow us to choose an optimal  weight, leading to a 
remarkable improvement of the bounds on the spacelike axis.

In Fig. \ref{fig:fig4} we demonstrate the effect of an additional input from the  
spacelike datum (\ref{eq:Huber2}),  using for illustration the weight $\rho_{1/2}(t)$.  
The solid lines show the bounds obtained as before without any spacelike information, 
while the dashed lines are obtained by imposing the central value from (\ref{eq:Huber2}). 
The inclusion of this additional information narrows considerably the allowed domain 
situated between the upper and the lower bounds.

The comparison of various weights is seen in Fig. \ref{fig:weights}, where the bounds are 
again obtained with the central input values, including the spacelike datum from (\ref{eq:Huber2}). 
At low $Q^2$ the bounds obtained with different  weights are almost identical, and are very tight. 
At higher $Q^2$ the small constraining effect of the weight $\rho_2(t)$ is again visible, while the 
other weights give  comparable results.  By inspecting the curves, we conclude that the best results 
are obtained with the weight $\rho_{1/2}(t)$, and we shall adopt this weight in what follows. In fact, 
it gives bounds almost identical with the weight  $\rho_1(t)\equiv 1$. In addition, it has the advantage 
of being less sensitive to the unknown high-energy behavior of the form factor, as discussed in the 
previous section in connection with the calculation of the quantities $I$ given in Table \ref{table:Ia}. 

\begin{figure}[htb]
 \vspace{0.6cm}
\begin{center}
  \includegraphics[width = 8.cm]{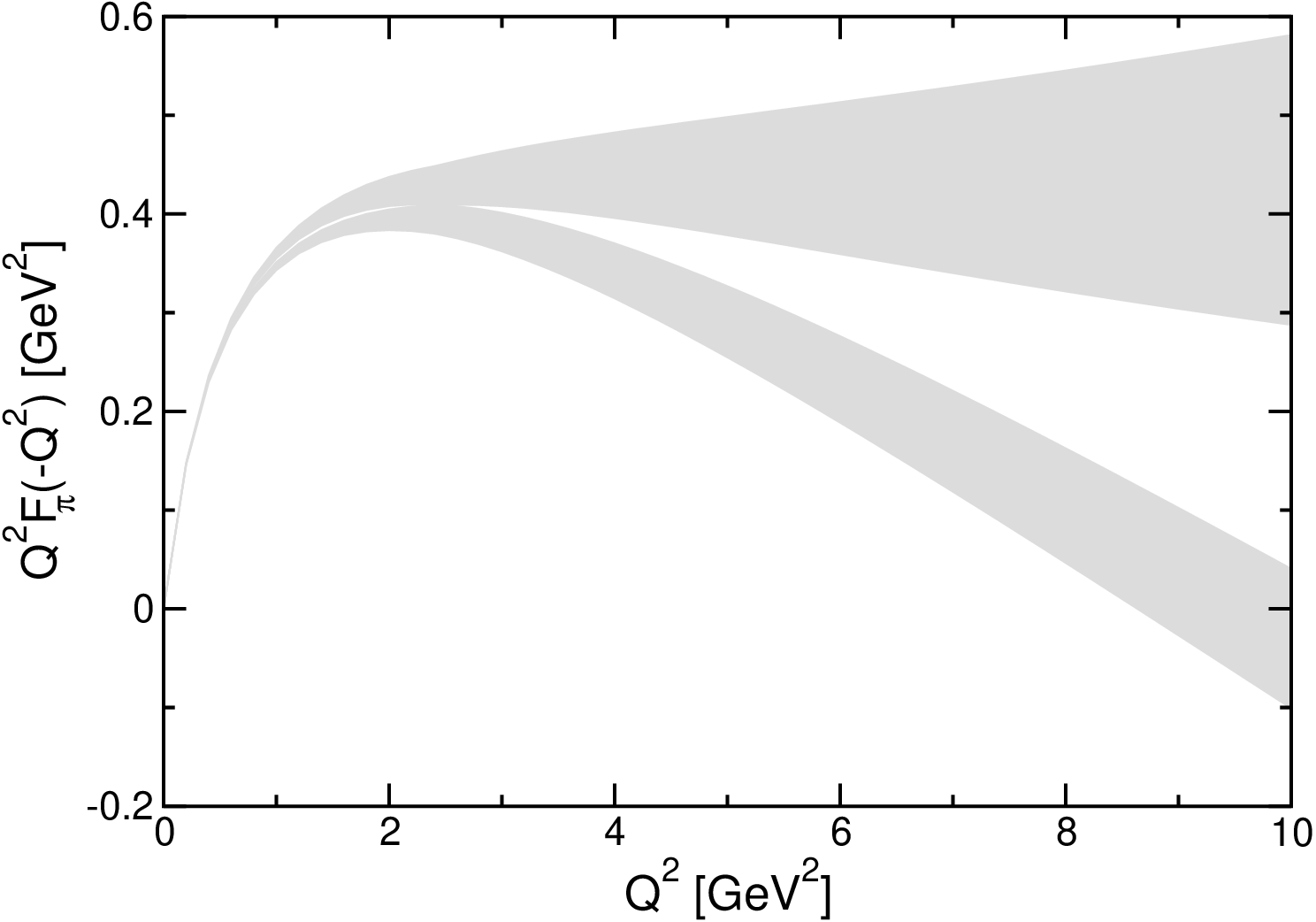}
 \caption{Upper and lower bounds derived with the weight $\rho_{1/2}(t)$ by including the charge radius (\ref{eq:r2}) and the spacelike datum (\ref{eq:Huber2}). Inner white region: allowed domain for the central values of the input. 
Grey bands: enlarged allowed domain with inclusion of the errors of the input. }
 \label{fig:fig5}
 	\end{center}
 \end{figure}

In Fig. \ref{fig:fig5} we illustrate, for the special weight $\rho_{1/2}(t)$, the effect of the 
uncertainties of the input on the resulting bounds. Since we are  interested in the most conservative 
results, {\em i.e.,} the largest allowed domain  defined by the upper and the lower bounds, it is enough 
to exhibit the upward shift of the upper bound, and the downward shift of  the lower bound, produced by 
the uncertainties of the input. By including these uncertainties,  the allowed domain obtained with 
the central input, shown as the inner white region in Fig.\ref{fig:fig5},  is enlarged covering the 
large grey domains.   We recall that in each point the error is obtained by adding quadratically 
the errors produced by the variation of the phase (\ref{eq:delta11}), the charge radius  (\ref{eq:r2}), 
the integral $I$ for $a=1/2$ from Table \ref{table:Ia}, and the experimental value given in (\ref{eq:Huber2}). 

It turns out that the greatest contribution to the size of the grey domain is the experimental 
uncertainty of the spacelike value  (\ref{eq:Huber2}), which decreases the lower bound by 
about 30\% for values of $Q^2$ around $7 \gev^2$, for instance. This shows that more accurate 
experimental data or lattice calculations at some points on the spacelike axis will lead to
 more stringent constraints  in this region, which, as shown below, is of interest for the onset 
of the perturbative QCD regime. At larger values of $Q^2$, the effect of the uncertainty in the 
charge radius starts to increase, becoming of comparable value, of about $30\%$, near $Q^2=10 \gev^2$. 
The uncertainty of the integral $I$ has an effect of less than 8\% in the considered range (as expected,
it has a noticeable effect only at higher momenta). 

\begin{figure}[htb]
 \vspace{0.35cm}
\begin{center}
  \includegraphics[width = 8.cm]{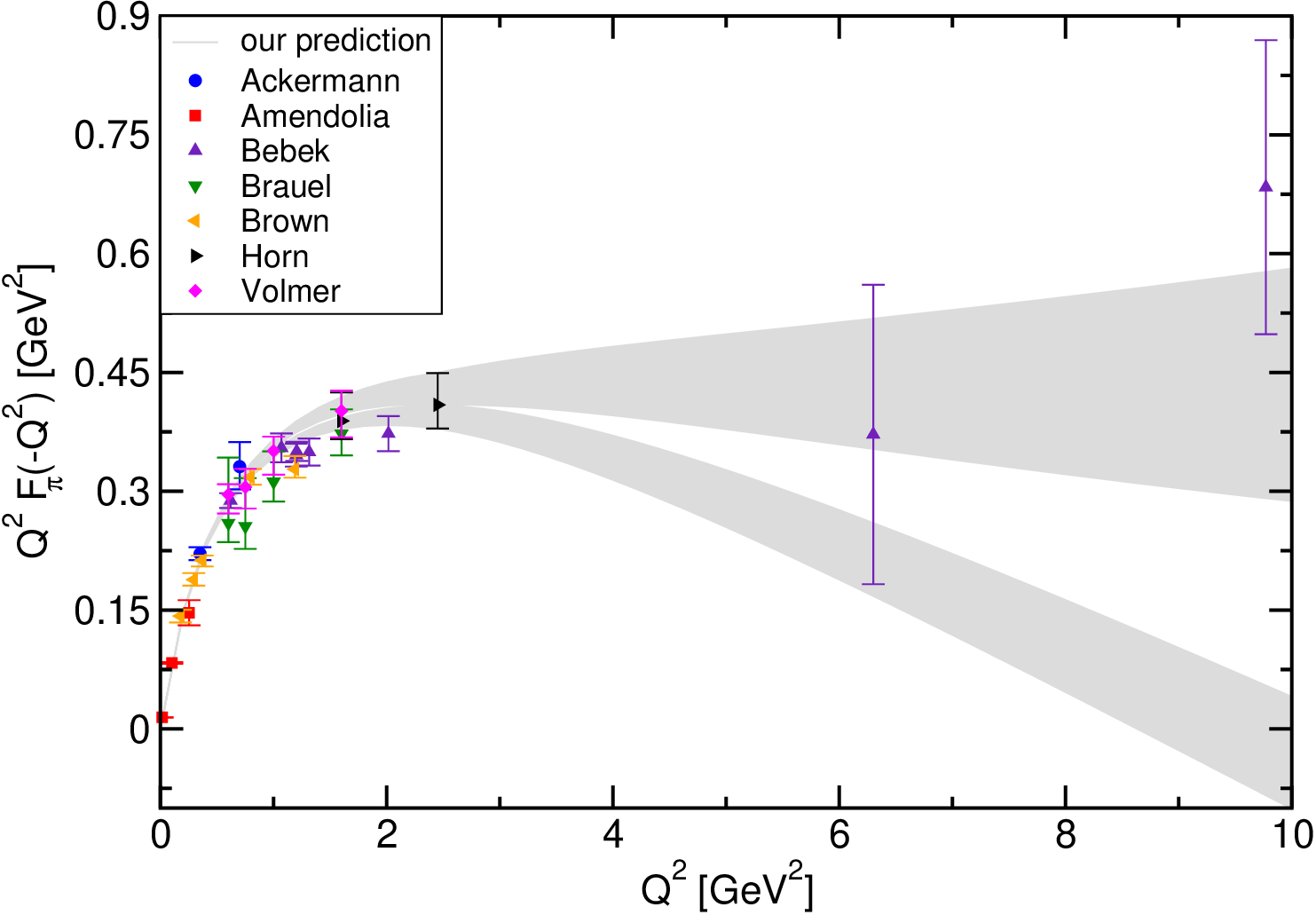}
 \caption{Allowed domain obtained with the weight $\rho_{1/2}(t)$  compared with several sets of experimental data.}
 \label{fig:fig6}
 	\end{center}
 \end{figure}

\begin{figure}[htb]
 \vspace{0.35cm}
\begin{center}
  \includegraphics[width = 8.cm]{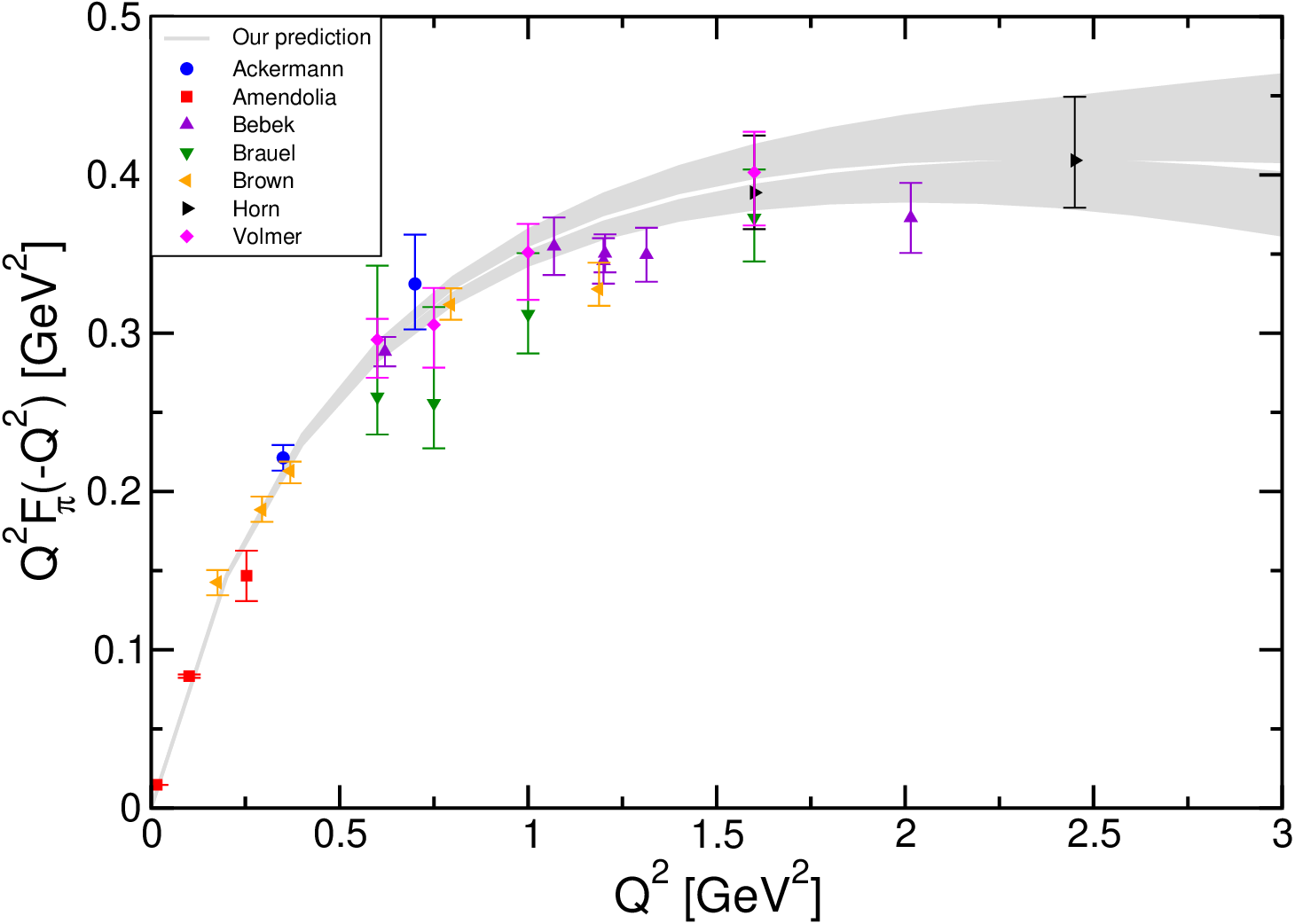}
 \caption{Enlarged view of Fig. \ref{fig:fig5} at low energy.}
 \label{fig:fig7}
 	\end{center}
 \end{figure}

In Figs.  \ref{fig:fig6} and  \ref{fig:fig7},  we compare the constraints corresponding to the 
weight $\rho_{1/2}(t)$ with some of the data available from experiments 
\cite{Horn,Huber,Volmer:2000ek,Ackermann:1977rp,Brauel:1979zk, Bebek:1974iz,Bebek:1976ww,Brown:1973wr,Amendolia:1986wj}. 
We find that at low $Q^2$ most of the low-energy data of Amendolia \cite{Amendolia:1986wj} are 
consistent with the narrow 
allowed band predicted by our analysis. The most recent data from \cite{Horn, Huber}
are well accommodated within our band, which is not surprising as we used one of these points as input. 
There are however some inconsistencies between the allowed domain derived here and the data of 
Bebek et al  \cite{Bebek:1974iz,Bebek:1976ww}, Ackermann et al \cite{Ackermann:1977rp}, and Brauel et al 
\cite{Brauel:1979zk}, in spite of their rather large errors.

\bigskip

\begin{figure}[htb]
 \vspace{0.35cm}
\begin{center}
  \includegraphics[width = 8.cm]{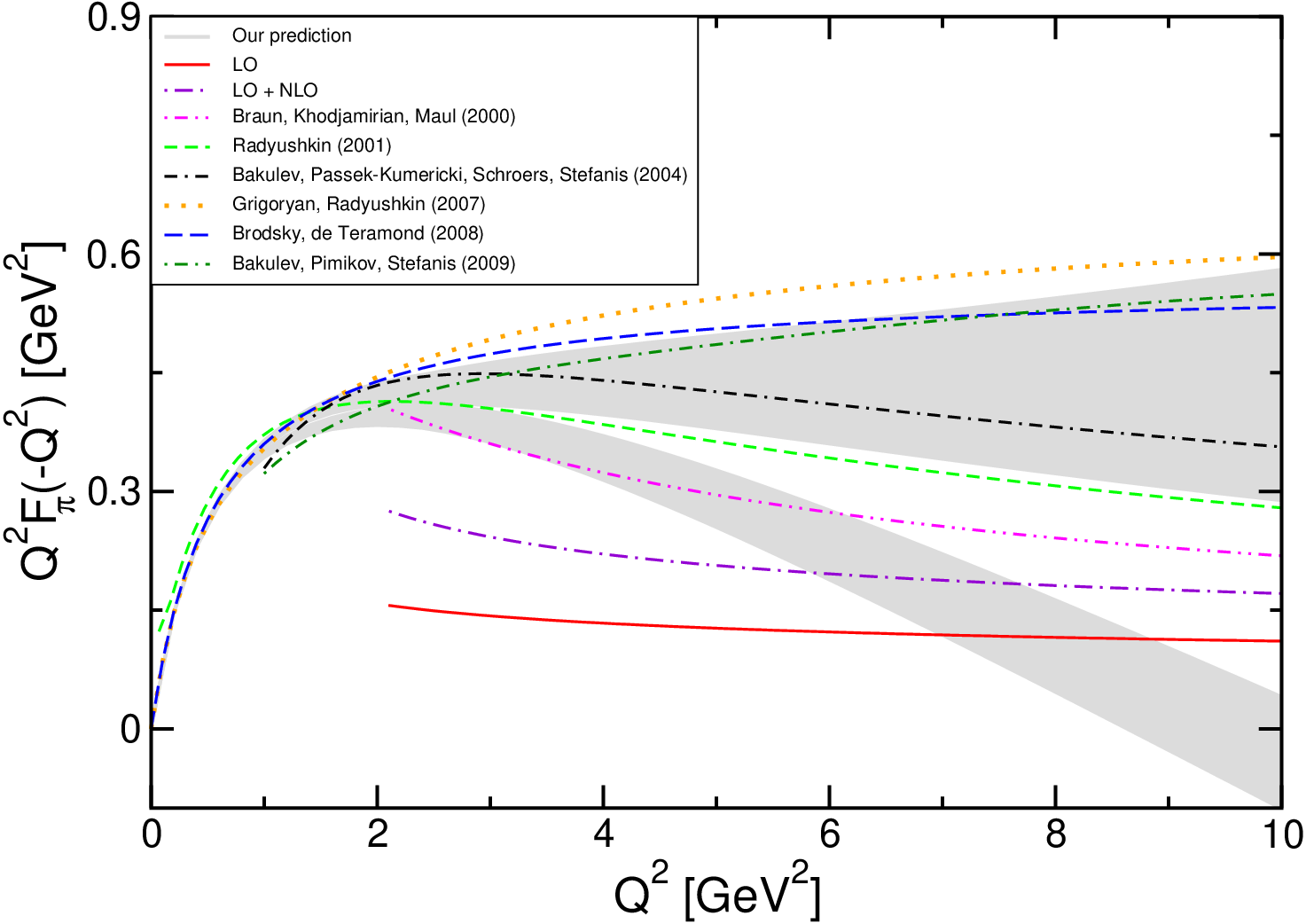}
 \caption{Comparison of the bounds with perturbative QCD and several nonperturbative models.}
 \label{fig:fig8}
 	\end{center}
 \end{figure}

Finally, in Fig. \ref{fig:fig8} we compare the allowed domain obtained in this work with the 
predictions of perturbative QCD and several nonperturbative models. We show first the LO 
expression (\ref{eq:qcd1}), in which we have taken the scale $\mu^2=Q^2$ and used the 
running coupling to one loop 
\begin{equation}\label{alphaeqn}
\alpha_s(Q^2)=\frac{4 \pi }{\beta_0 \ln(Q^2/\Lambda^2)}.
\end{equation}
We have taken $n_f=3$ active flavors and $\Lambda=0.214\, \gev$, which
in Eq. (\ref{alphaeqn}) gives  $\alpha_s(M_\tau^2)=0.33$,  the average of the various predictions
from hadronic $\tau$ decays obtained recently  
\cite{Davier:2008,BJ,CaFi:2009,CaFi:2011,Muenchen,Boito:2011,Abbas:2012}. 
As seen from   Fig. \ref{fig:fig8}, the central LO curve, obtained with this choice of the coupling,  stays below the lower bound derived in the 
present work up to $Q^2= 7\, \gev^2$. The  variations  set by 
the error of about $\pm 0.02$ in $\alpha_s(M_\tau^2)$ quoted in \cite{Davier:2008,BJ,CaFi:2009,CaFi:2011,Muenchen,Boito:2011,Abbas:2012}
do not modify this conclusion.

As discussed in  \cite{Brodsky:1998, Melic:1999mx} the perturbative prediction to NLO  
is sensitive to the choice of the renormalization scale, and also of the factorization scale in the case 
when pion DAs different from the asymptotic ones are used in the calculation. 
Several prescriptions for scale setting were adopted, but there is no general consensus on the issue. 
In particular, in the Brodsky-Lepage-Mackenzie procedure the renormalization scale is chosen  
such that the coefficient of $\beta_0$ in the NLO correction (\ref{eq:qcd2}) vanishes \cite{Brodsky:1998}.

 For illustration, in  Fig. \ref{fig:fig8}  we show the sum of the LO and NLO terms  (\ref{eq:qcd1}) and 
(\ref{eq:qcd2}),  obtained with the  scale $\mu^2=Q^2$ and the one loop coupling (\ref{alphaeqn}).   As in previous references \cite{DoNa, Melic:1998qr, Melic:1999mx, Braun:1999uj,RuizArriola},  we have taken the number of active flavors $n_f=3$. This curve is compatible with our bounds enlarged by errors only for $Q^2>6\, \gev^2$. 

When $Q^2$ is varied over a large range, one must of course take into account the transition from $n_f$ to $n_f+1$ active flavors in the $\beta$-function coefficients when the threshold of a new flavor is crossed.  However, it turns out that in the region of interest to us, $Q^2<10\, \gev^2$, the effect of the higher mass quarks is small.  First, we recall that the definition of the flavor matching scale is subject to a certain uncertainty. If the thresholds are set at the pole masses of the charm and bottom quarks,  $n_f$ should be raised to 4 and 5, respectively,
when the  thresholds of 1.65 GeV and 4.75 GeV are crossed.  However, the change in $\beta_0$ is small, from $\beta_0=9$ to $\beta_0=8.33$ for the charm threshold relevant in the region considered.  Moreover, the continuity  of the coupling (\ref{alphaeqn}) must be imposed by a change of the parameter $\Lambda$ above the threshold. Therefore, the modification of the form factor below $Q^2=10\, \gev^2$ is expected to be quite small and the comparison with the bounds derived in the present paper remains unchanged. We note also that for the choice of the quark-flavor matching scales at $2 m_q$  as in \cite{Davier:2008}, or at $\mu_c^*=3.729\, \gev$ and $\mu_b^*=10.588\, \gev$ as in \cite{BJ}, the
thresholds are not crossed and the number of the active flavors remains equal to three in the whole region shown in  Fig. \ref{fig:fig8}.  

 In Fig.  \ref{fig:fig8} we show also several  nonperturbative models proposed in the literature 
for the spacelike form factor at intermediate region
\cite{Braun:1999uj, Radyushkin:2001,Bakulev:2004cu, Brodsky:2007hb,Grigoryan:2007, Braguta:2007fj, Bakulev:2009ib}

 In Ref. \cite{Braun:1999uj}, the authors applied light-cone QCD sum rules and parametrized with a 
simple expression the nonperturbative  correction, to be added to the LO+NLO perturbative prediction 
in the region $1<Q^2<15\,\gev^2$.  In Fig. \ref{fig:fig8}  we show the sum of the soft correction and the 
perturbative QCD prediction to NLO,  evaluated at a scale $\mu^2=0.5\, Q^2 + M^2$ with $M^2=1\,\gev^2$ 
as argued in \cite{Braun:1999uj}. The model is quite compatible with our bounds, the corresponding curve
being inside the small white inner domain for $Q^2>6\, \gev^2$.

 The model based on local duality \cite{Radyushkin:2001} is also consistent with the allowed 
domain derived here for $Q^2> 1\,\gev^2$. We mention that this model, proposed in \cite{Nesterenko:1982}, was 
recently developed by several authors  \cite{Braguta:2007fj}.  The other models shown in  Fig. 
\ref{fig:fig8} are consistent with the bounds derived by us at low $Q^2$, but are at the upper limit 
of the allowed domain at higher $Q^2$. The agreement is somewhat better for the model discussed in 
\cite{Bakulev:2004cu}, which is a LO+NLO perturbative calculation using nonasymptotic pion DAs evolved 
to NLO, with a modification of the QCD coupling by the so-called analytic perturbation theory. 
The AdS/QCD model considered in \cite{Brodsky:2007hb} is in fact a simple dipole interpolation, 
which is valid at low energies but seems to overestimate the form factor
at larger momenta. The same remark holds for the models 
discussed in  \cite{Bakulev:2009ib} and \cite{Grigoryan:2007},  based on QCD sum rules with nonlocal condensates, and the chiral limit of the 
hard-wall AdS/QCD approach, respectively.

\bigskip

\section{Discussion and Conclusions}\label{sec:conclusions}
In this paper we derived upper and lower bounds on the pion electromagnetic form factor along the spacelike 
axis, by exploiting in a conservative way the precise recent information on the phase and modulus on 
complementary regions of the timelike axis. 

More exactly, we have applied a method of analytic continuation which uses as input the phase only in the 
elastic region $t<\tin$, where it is known with high accuracy from the dispersive theory of $\pi\pi$ scattering 
via the Fermi-Watson theorem. It can be shown rigorously \cite{Abbas:2010EPJA} that the results are 
independent of the unknown phase of the form factor above the first inelastic threshold  $\tin$. As 
for the modulus, we have used the BaBar data in the range between the  first inelastic threshold and 3 GeV \cite{BABAR}, 
and very conservative assumptions  above 3 GeV. The results are not very sensitive to 
these assumptions, since we include the information on the modulus through the $L^2$ integral condition 
(\ref{eq:L2}), instead of imposing this knowledge pointwise, at each $t$. Obviously,  the detailed 
behavior of $|F(t)|$ is averaged in the integral, and the high-energy contribution can be suppressed 
 by a suitable choice of the weight $\rho(t)$.
In our calculations we have choosen a suitable weight, which on the one hand gives sufficiently  tight bounds,  
and on the other hand ensures a small sensitivity to the high energy part. We have found that these constraints 
are met by a weight of the simple form  (\ref{eq:rhogeneric}), with the choice $a=1/2$. 

We mention that, for completeness, we have investigated also the more general class of weight 
functions of the form  (\ref{eq:rhogeneric0}) and found, for instance, that with the choices 
$b=1/2$, $c=1$ and $Q_0^2$ in the range $3-6\,\gev^2$ the bounds are slightly better and have 
smaller uncertainties than those shown in Fig. \ref{fig:fig5}. For the 
conclusions formulated in this paper the improvement is not essential.  However,  a more 
systematic optimization  with respect to the weight  is of interest for further studies, 
especially when more accurate input from the spacelike axis will be available.

Besides the use of the modulus above $\tin$ only in an averaged way, as in the condition 
(\ref{eq:L2}), we note that our method does not exploit completely the present knowledge 
of the modulus, since we do not use the data on  $|F(t)|$ below $\tin$. While it may seem 
that this additional information can improve in a significant way the bounds, it turns out 
that it is not so. Indeed, the knowledge of the phase below $\tin$ has, in the case of the 
pion form factor, a considerable  constraining power on the modulus in the same region. The 
reason is the strong peak produced by the $\rho$ resonance.  The Omn\`es function $\omnes(t)$ 
defined in (\ref{eq:omnes}) reproduces well this impressive behavior of the modulus, and this 
holds for a large class of choices of the arbitrary phase above the elastic region. The effect 
of this unknown  phase is actually completely removed in our formalism by the information on the 
modulus, which constrains the auxiliary function $h(t)$ appearing in the expression (\ref{eq:h}) 
of $F(t)$.  Therefore, imposing additional data on  $|F(t)|$ below 0.9 GeV is not expected to 
improve considerably the bounds, if the input values of the phase and modulus in this region are 
consistent. Of course, one can reverse the argument and find constraints on the modulus on the 
timelike axis below $\tin$, in order to test the consistency. This analysis will be carried out
in the future.

 In the present work we have investigated the consistency of the timelike phase 
and modulus with the experimental data and theoretical models on the spacelike axis. The slow 
onset of the asymptotic regime predicted by factorization and perturbative QCD for the pion 
electromagnetic form factor has been known for a long time.  A lot of theoretical work has been done  
by several groups on the determination of the pion  form factor  in the spacelike intermediate
 energy region, where the soft, nonperturbative contributions are expected to be important. 
However, definite conclusions on the validity of the theoretical models and the precise energy 
at which the asymptotic regime may be considered valid  were not possible, due to the lack of 
accurate experimental data in the relevant region. 

The bounds derived in the present paper are almost model-independent and very robust, allowing 
us to make some definite statements. From Fig. \ref{fig:fig8} it is possible to  say with great 
confidence that perturbative QCD to LO is excluded  for $Q^2< 7\,\gev^2$,  and perturbative QCD 
to NLO is excluded  for $Q^2< 6\,\gev^2$, respectively. If we restrict to the inner white allowed 
domain obtained with the central values of the input, the exclusion regions  become 
$Q^2< 9\,\gev^2$ and  $Q^2< 8\,\gev^2$, respectively. Among the theoretical models, the 
light-cone QCD sum rules \cite{Braun:1999uj} and the local quark-hadron duality model \cite{Radyushkin:2001}  
are consistent with the allowed domain derived here for a large energy interval, while the remaining 
models are consistent with the bounds at low energies, but seem to predict too high values at higher $Q^2$.

To increase the strength of the predictions, a reduction of the 
grey bands produced by the uncertainties of the input is desirable. 
As we mentioned, the biggest contribution is brought 
by the experimental errors of the input spacelike datum (\ref{eq:Huber2}).  
Therefore, more accurate data at a few spacelike points, particularly
at larger values of $Q^2$ will be very important for increasing the 
predictive power of the formalism developed and applied in the present paper.

\vskip0.2cm
\noindent{\bf Acknowledgement:} 
 IC acknowledges support from  CNCS in the Program Idei, Contract No.
121/2011. We thank B. Malaescu for correspondence on the BaBar data
and S.Dubni$\check{c}$ka for discussions.

\end{document}